\documentclass[10pt,conference]{IEEEtran}
\IEEEoverridecommandlockouts
\usepackage{cite}
\usepackage{amsmath,amssymb,amsfonts}
\usepackage{algorithmic}
\usepackage{graphicx}
\usepackage{textcomp}
\usepackage{xcolor}
\usepackage{float}
\usepackage{enumerate}
 \usepackage{multirow}
\usepackage{epstopdf}
\usepackage{mathtools, cuted}
\allowdisplaybreaks
\usepackage[-5, nonofiles]{pagesel} 
\def\BibTeX{{\rm B\kern-.05em{\sc i\kern-.025em b}\kern-.08em
    T\kern-.1667em\lower.7ex\hbox{E}\kern-.125emX}}
\begin{document}
\title{BER Analysis of RIS Assisted Multicast Communications with Network Coding
}

\author{\IEEEauthorblockN{Vetrivel Chelian Thirumavalavan}
\IEEEauthorblockA{\textit{Research Scholar, Dept. of ECE} \\
\textit{Thiagarajar College of Engineering}\\
Madurai, India \\
vetrivelchelian@student.tce.edu}
\and
\IEEEauthorblockN{PGS Velmurugan}
\IEEEauthorblockA{\textit{Assistant Professor, Dept. of ECE} \\
\textit{Thiagarajar College of Engineering}\\
Madurai, India \\
pgsvels@tce.edu}

\and
\IEEEauthorblockN{Thiruvengadam S J}
\IEEEauthorblockA{\textit{Professor, Dept. of ECE} \\
\textit{Thiagarajar College of Engineering}\\
Madurai, India \\
sjtece@tce.edu}

}
\maketitle
\begin{abstract}
In this paper, Reconﬁgurable Intelligent Surface (RIS) assisted dual-hop multicast wireless communication network is proposed with two source nodes and two destination nodes. RIS boosts received signal strength through an intelligent software-controlled array of discrete phase-shifting metamaterials. The multicast communication from the source nodes is enabled using a Decode and Forward (DF) relay node. In the relay node, the Physical Layer Network Coding (PLNC) concept is applied and the PLNC symbol is transmitted to the destination nodes. The joint RIS-Multicast channels between source nodes and the relay node are modeled as the sum of two scaled non-central Chi-Square distributions. Analytical expressions are derived for Bit Error Rate (BER) at relay node and destination nodes using Moment Generating Function (MGF) approach and the results are validated using Monte-Carlo simulations. It is observed that the BER performance of the proposed RIS assisted network is a lot better than the conventional non-RIS channels links. 
\end{abstract}


\section{Introduction }
RIS is a new frontier in wireless communications to improve reliability of the system and it has been widely investigated in theory for the past couple of years \cite{di2019smart}. The key idea behind the invention of RIS was the introduction of tunable meta-surfaces \cite{kaina2014shaping}. In a very recent study, it is reported the RIS systems' prototyping is successful when compared with the conventional phased array systems, RIS systems are ultra-energy efficient and is validated experimentally \cite{dai2019reconfigurable}. Theoretical propagation and pathloss modeling of RIS are reported in \cite{zdogan2019intelligent}, experimental support for the theoretical pathloss model for RIS based communications is proven in\cite{tang2019wireless}. Orthogonal Frequency Division Multiplexing (OFDM) and passive beamforming case for RIS based communications are investigated in \cite{yang2019irs,yang2019intelligent}. 

In multicast communication, a single source node sends data to multiple receivers by exploiting the broadcast nature of channels. It also hugely improves group spectral efficiency \cite{lee2016ber}. Multicast communications have huge potential in future Machine type-Internet of Things (Mt-IoT) mainly to simultaneously send control messages towards a huge number of IoT devices i.e., group paging\cite{6476030}.
\begin{figure}
\centering
\includegraphics[width=\linewidth]{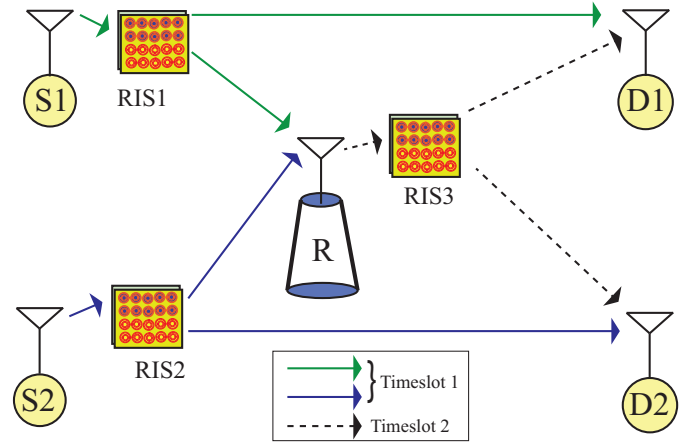}
\caption{Proposed RIS assisted Multicast Communications with Network Coding}
\label{Figure_1}
\end{figure}

Physical Layer Network Coding (PLNC) is a technique where interference between bitstreams becomes arithmetic operations in network coding directly within the radio channel at the physical layer, Employing PLNC also reduces overall latency \cite{liew2013physical}. Outage performance of multicast full-duplex cognitive radio system with PLNC is investigated for Nakagami-m fading environment in \cite{velmurugan2015full}.

The major contributions of this paper are
\begin{itemize}
\item RIS based communication is extended to multicast system with two source nodes and two destination nodes with a PLNC based DF relay node.
\item Exact closed-form analytical BER expressions are derived for the RIS assisted multicast communication at relay node in fading environment and the end to end BER performance is analyzed using Monte-Carlo computer simulations.
\end{itemize}

\section{System Model}

The proposed RIS assisted multicast communication network is shown in Figure \ref{Figure_1}. In the proposed network, RISs are employed at both source nodes $S_1$ $\&$ $S_2$ and relay node $R$ in close proximity. The passive reflecting elements of RIS boost the received signal strengths at relay node $R$, destination nodes $D_1$ and $D_2$. Let $N_1$, $N_2$ and $N_3$ be the total number of reflecting elements at RIS1, RIS2 and RIS3 respectively. The source nodes $S_1$ and $S_2$ transmit their symbols to both the destination nodes $D_1$ and $D_2$. However, the cross links $S_1$ to $D_2$ and $S_2$ to $D_1$ are not available due to large scale path loss. Hence, DF relay node is used to transmit symbols from $S_1$ to $D_2$ and $S_2$ to $D_1$ using network coding concept.

\subsection{Time Slot - 1}
At time slot 1, the received signals at $D_1$ and $D_2$ are written as
\begin{equation}
y_{S_1D_1}=\left[ \sqrt{Ps_1} \mathbf{h}_{S_1D_1}\Phi_1 \right] x_1 + n_{D_1} 
\label{eqn1}
\end{equation}
\begin{equation}
y_{S_2D_2}=\left[ \sqrt{Ps_2}\mathbf{h}_{S_2D_2}\Psi_1 \right] x_2 + n_{D_2} 
\label{eqn2}
\end{equation}
$Ps_1$ and $Ps_2$ are the transmitted signal powers at $S_1$ and $S_2$ respectively. $x_1$ and $x_2$ are the transmitted symbols from $S_1$ and $S_2$ respectively. $\mathbf h_{S_1D_1}$ is $(1\times\frac{N_1}{2})$ fading channel coefficient vector between $S_1$ and $D_1$. $\mathbf h_{S_2D_2}$ is $(1\times\frac{N_2}{2})$ fading channel coefficient vector between $S_2$ and $D_2$. The channel coefficents are modeled as independent zero mean circularly symmetric complex Gaussian (ZMCSCG) random variables. $\mathbf h_{S_1D_1}$ and $\mathbf h_{S_2D_2}$ are defined as
\begin{equation*}
\mathbf{h}_{S_1D_1}=\left[\alpha_1 e^{-\theta_1}, \ldots ,\alpha_i e^{-\theta_i}, \ldots \alpha_\frac{{N_1}}{2} e^{-\theta_\frac{{N_1}}{2}} \right]
\end{equation*}
\begin{equation*}
 \mathbf{h}_{S_2D_2}=\left[\beta_1 e^{-\psi_1}, \ldots ,\beta_k e^{-\psi_k}, \ldots \beta_\frac{{N_2}}{2} e^{-\psi_\frac{{N_2}}{2}} \right]
 \end{equation*}
$\alpha_i$ and $\beta_k$ are the magnitude components of the $i^{th}$ and $k^{th}$ channel coefficients $\mathbf{h}_{S_1D_1}$ and $\mathbf{h}_{S_2D_2}$ respectively. $\alpha_i$ and $\beta_k$ are Rayleigh distributed with $E[\alpha_i]=E[\beta_k]=\sqrt{\frac{\pi}{4}}$ \& $VAR[\alpha_i]=VAR[\beta_k]=\frac{4-\pi}{4}$ \cite{salo2006distribution}. RIS1 phasing is denoted as 
\begin{equation*}
\Phi_1= \eta_1 \left[e^{j\phi_1},\ldots, e^{j\phi_i}, \ldots, e^{j\phi_\frac{{N_1}}{2}} \right]^T
\end{equation*}
and RIS2 phasing is expressed as 
\begin{equation*}
\Psi_1= \eta_2 \left[e^{j\varphi_1},\ldots, e^{j\varphi_k}, \ldots, e^{j\varphi_\frac{{N_2}}{2}} \right]^T, 
\end{equation*}
$\eta_1,\eta_2 \in (0,1]$ are the reflection loss coefficients of RIS1 and RIS2 respectively. $\phi_i$ and $\psi_k$ are the phase shifts introduced by the $i^{th}$ reflecting element of the RIS1 and $k^{th}$ reflecting element of the RIS2 respectively. $n_{D_1}$ and $n_{D_2}$ are ZMCSCG white noise with variance $N_0$.

Relay node receives signals from both $S_1$ and $S_2$ at the same time slot. Hence, the received signal at relay node R is given by
\begin{equation}
y_{R}=\left[ \sqrt{Ps_1}\mathbf{h}_{S_1R}\Phi_2\right] x_1+\left[ \sqrt{Ps_2}\mathbf{h}_{S_2R}\Psi_2 \right] x_2+n_{R} 
\end{equation}
 $\mathbf{h}_{S_1R}$ and $\mathbf{h}_{S_2R}$ are the fading channel coefficient vector between RIS1 and relay node $R$ \& RIS2 and relay node $R$ respectively, and is defined as
 \begin{equation*}
 \mathbf{h}_{S_1R}=\left[\alpha_{\frac{{N_1}}{2}+1} e^{-\theta_{\frac{N_{1}}{2}+1}}, \ldots ,\alpha_i e^{-\theta_i}, \ldots \alpha_{N_1} e^{-\theta_{N_1}} \right]^T
 \end{equation*}
 \begin{equation*}
 \mathbf{h}_{S_2R}=\left[\beta_{\frac{{N_2}}{2}+1} e^{-\psi_{\frac{N_{2}}{2}+1}}, \ldots ,\beta_k e^{-\psi_k}, \ldots \beta_{N_2} e^{-\psi_{N_2}} \right]^T
 \end{equation*}
$n_R$ is ZMCSCG white noise with variance $N_0$ at relay node $R$. Similarly, phasing of RIS1 and RIS2 are denoted as 
\begin{equation*}
\Phi_2= \eta_1 \left[e^{j\phi_{\frac{{N_1}}{2}+1}},\ldots, e^{j\phi_i}, \ldots, e^{j\phi_{{N_1}}} \right]
\end{equation*}
\begin{equation*}
\Psi_2= \eta_2 \left[e^{j\varphi_{\frac{{N_2}}{2}+1}},\ldots, e^{j\varphi_k}, \ldots, e^{j\varphi_{{N_2}}} \right]
\end{equation*}

\subsection{Time Slot - 2}
 Let $ \widetilde{x} = \left[\hat{x}_{1} \oplus \hat{x}_{2}\right]$ and at time slot 2, the relay broadcasts the data $\widetilde{x}$ to $D_1$ and $D_2$. $D_1$ already has $x_1$, therefore PLNC data from $D_2$ $(x_2)$ is obtained by XOR operation of $x_1$ with $\widetilde{x}$. Similarly, $x_1=x_2 \oplus \widetilde{x}$. 
\begin{figure}
\centering
\includegraphics[width=\linewidth]{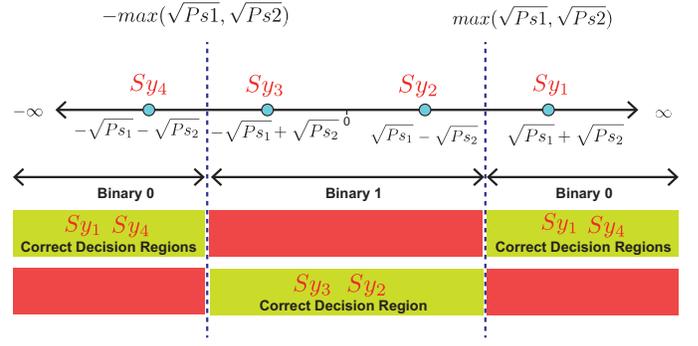}
\caption{Received Signal Constellation at Relay Node R. Here $Ps_1$ > $Ps_2$}
\label{const}
\end{figure}

\begin{equation}
y_{RD_1}=\sqrt{P_R}\left[\mathbf{h}_{RD_1} \xi_1 \right] \widetilde{x} +n_{D_1}
\end{equation}
\begin{equation}
y_{RD_2}=\sqrt{P_R}\left[\mathbf{h}_{RD_2} \xi_2 \right] \widetilde{x} +n_{D_2}
\end{equation}
$\left[\mathbf{h}_{RD_1} \xi_1 \right] $ and $\left[\mathbf{h}_{RD_2} \xi_2 \right]$ are the RIS channels between relay to destinations $D_1$ and $D_2$.

\section{Performance Analysis}

Considering BPSK signaling from both $S_1$ and $S_2$, relay can receive four possible superimposed symbols $\{Sy_1, Sy_2, Sy_3, Sy_4\}=\{(\sqrt{Ps_1} + \sqrt{Ps_2}),~ (\sqrt{Ps_1} - \sqrt{Ps_2}),~ (- \sqrt{Ps_1} + \sqrt{Ps_2}),~ (-\sqrt{Ps_1} - \sqrt{Ps_2}) \}$. Fig. \ref{const} shows the received BPSK symbols from two source nodes ($S_1$ and $S_2$) at the relay node and it also describes the correct regions for the symbols.

Received signal at relay in an AWGN environment is written as 
\begin{equation}
y_{R}=\sqrt{Ps_1}x_1+\sqrt{Ps_2} x_2+n_{R}
\end{equation}
and so the Signal to Noise Ratio (SNR) is written as 
\begin{equation}
\gamma_{R}^{AWGN}=\frac{\left(Ps_1\right)+\left({Ps_2}\right)}{2N_0}
\label{awgnSNR}
\end{equation}

PLNC based detection is employed in the relay node $R$, the detection is similar to duo-binary decoding except that $Ps_1$ and $Ps_2$ are different. In AWGN environment, ${Sy_4}, {Sy_1}$ are same sign pairs are the extreme end constellation points. The decisions for end constellation points are $binary~0$. The decisions for the alternate sign pair intermediate points $({Sy_2}, {Sy_3})$ are taken as $binary~1$. The average error probabilities for the aforementioned cases are derived as
\begin{multline}
Pe^{AWGN}, Sy_{1} = Pe^{AWGN}, Sy_{4}=\\ \frac{1}{\sqrt{2\pi N_{0}}} \int_{-max(\sqrt{Ps_1},\sqrt{Ps_2})}^{max(\sqrt{Ps_1},\sqrt{Ps_2})} e^{\frac{-\left(y-\left(\sqrt{Ps_1} + \sqrt{Ps_2}\right)\right)^2}{2N0}} dy
\end{multline}
but for symbols $Sy_2$ and $Sy_3$ the error regions are $-\infty$ to $-\sqrt{Ps1}$ and $\sqrt{Ps2}$ to $-\infty$,
\begin{multline}
Pe^{AWGN}, Sy_{2} = Pe^{AWGN}, Sy_{3}=\\ 1-\frac{1}{\sqrt{2\pi N_{0}}} \int_{-max(\sqrt{Ps_1},\sqrt{Ps_2})}^{max(\sqrt{Ps_1},\sqrt{Ps_2})} e^{\frac{-\left(y-\left(+\sqrt{Ps_1} - \sqrt{Ps_2}\right)\right)^2}{2N0}} dy 
\end{multline}
Since all symbols are equiprobable, The exact average probability of error for AWGN environment channels is given by 
\begin{multline}
Pe= Q\left(\sqrt{{\frac{ ~min(Ps_1, Ps_2)}{N_0}}}\right)- \\ \frac{1}{2}Q\left(\frac{2\sqrt{max(Ps_1, Ps_2)}+\sqrt{ ~min(Ps_1, Ps_2)}}{\sqrt{N_0}}\right) \\ +\frac{1}{2}Q\left(\frac{2\sqrt{max(Ps_1, Ps_2)}-\sqrt{ ~min(Ps_1, Ps_2)}}{\sqrt{N_0}}\right)
\label{awgnexact1}
\end{multline} 
At high SNR regions only the pairs with minimum Euclidean distances to thresholds contribute most errors, therefore (\ref{awgnexact1}) can be approximated as
\begin{equation}
Pe_{approx}^{AWGN}=Q\left(\sqrt{\frac{{ ~min(Ps_1, Ps_2)}}{N_0}}\right)
\label{AWGnapproxGen}
\end{equation}
From \cite{258319}, using Q-functions alternative Craig's form , i.e, $Q\left({x}\right)=\frac{1}{\pi} \int_{0}^{\frac{\pi}{2}} \exp \left(-{\left({x^2}\right)}/{2 \sin ^{2}\omega}\right) d  \omega$. (\ref{awgnexact1}) and (\ref{AWGnapproxGen}) are rewritten as
\begin{multline}
Pe_{exact}^{AWGN}= \frac{1}{\pi} \int_{0}^{\frac{\pi}{2}} \left. \exp \left(-\frac{ ~min(Ps_1, Ps_2)}{2N_0 \sin ^{2}  \omega}\right) \right.- \\ \exp \left(-\frac{\left({2\sqrt{max(Ps_1, Ps_2)}+\sqrt{ ~min(Ps_1, Ps_2)}}\right)^2}{2N_0 \sin ^{2}  \omega}\right) + \\ \left. \exp \left(-\frac{\left({2\sqrt{max(Ps_1, Ps_2)}-\sqrt{ ~min(Ps_1, Ps_2)}}\right)^2}{2N_0 \sin ^{2} \omega}\right)  \right. d  \omega
\label{exactpe}
\end{multline} 
\begin{equation}
Pe_{approx}^{AWGN}=\frac{1}{\pi} \int_{0}^{\frac{\pi}{2}} \exp \left(-\frac{ ~min(Ps_1, Ps_2)}{2N_0 \sin ^{2}  \omega}\right)  d \omega
\label{approxpe}
\end{equation}
Since the system resembles the uplink Non-Orthogonal Multiple Access (NOMA) scenario, using \cite{yeom2019ber}, the instantaneous SNR  $(\gamma_{R})$ at relay node R is given by,
\begin{equation}
\gamma_{R}=\frac{\left(\frac{Ps_1 \left|\mathbf{h}_{S_1R}\Phi_2 \right|^2}{N_0}\right)+\left(\frac{Ps_2 \left|\mathbf{h}_{S_2R}\Psi_2 \right|^2}{N_0}\right)}{2}
\end{equation}
Assuming ideal phase compensation at both RIS1 and RIS2, i.e, $\phi_i+\theta_i=0 ~\forall i$ and $i \in \{1,2,..,N_1 \}$, $\varphi_i+\psi_i=0 ~\forall k$ and $k\in \{1,2,..,N_2 \}$. The vector products $\mathbf{h}_{S_1R} \Phi_2$ and $\mathbf{h}_{S_2R} \Psi_2$ can be rewritten in summation form with magnitude components only. Hence, it is simplified as
\begin{equation}
\gamma_{R}= \frac{ Ps_1\left|\sum_{i=1}^{\frac{N_1}{2}}\alpha_i  \right|^2 + Ps_2\left|\sum_{k=1}^{\frac{N_2}{2}}\beta_i  \right|^2}{2 N_0}
\end{equation}
Let $\sum_{i=1}^{\frac{N_1}{2}}\alpha_i = \mathcal{A}$ and $\sum_{k=1}^{\frac{N_2}{2}}\beta_i= \mathcal{B}$. For large values of $N_1$ and $N_2$, according to central limit theorem,  $\mathcal{A}$ and $\mathcal{B}$ are approximated as Gaussian random variables with $\operatorname{E}[\mathcal{A}] =\frac{N_1\pi}{8}$ \& $\operatorname{VAR}[\mathcal{A}]=\frac{N_1}{2}\left(1-\frac{\pi^{2}}{16}\right)$ and similarly, $\operatorname{E}[\mathcal{B}]= \frac{N_2\pi}{8}$ \& $\operatorname{VAR}[\mathcal{B}]=\frac{N_2}{2}\left(1-\frac{\pi^{2}}{16}\right)$. Therefore $\mathcal{A}^2$, $\mathcal{B}^2$ converges to non-central Chi-Square distributed RVs, 
\begin{equation}
\gamma_{R}=\frac{\left(Ps_1\mathcal{A}^2\right)+\left({Ps_2\mathcal{B}^2}\right)}{2N_0}
\label{gammaR11}
\end{equation}
(\ref{gammaR11}) consists of a weighted sum of two non-central Chi-Square RVs. It is accurately modeled using Moment-Generating Function (MGF). By statistical properties, if a random variable $Y_{n}$ is defined as $Y_{n}=a_{1} X_{1} + a_{2} X_{2}$, where the $X_1 \& X_2$ are independent random variables and the $a_1 \& a_2$ are constants, the MGF of $Y_n$ is given by $M_{Y_{n}}(t)=M_{X_{1}}\left(a_{1} t\right) M_{X_{2}}\left(a_{2} t\right) $. Here, $Ps_1$ and $Ps_2$ are constants, Let $SNR_1=\frac{Ps_1}{N_0}$ and $SNR_2=\frac{Ps_2}{N_0}$. The joint MGF of $\gamma_{R}$ is given by
\begin{multline}
M_{\gamma_{R}}(s)=\\ \left({1-\frac{s N_1(4-\pi) SNR_1}{4}}\right)^{-0.5} \exp \left(\frac{\frac{-s N_1^{2} \pi SNR_1}{8}}{1-\frac{s N_1(4-\pi) SNR_1}{4}}\right) \times \\  \left({1-\frac{s N_2(4-\pi) SNR_2}{4}}\right)^{-0.5} \exp \left(\frac{\frac{-s N_2^{2} \pi SNR_2}{8}}{1-\frac{s N_2(4-\pi) SNR_2}{4}}\right)
\label{jointMGFd}
\end{multline}
The BER in fading environment is given by 
\begin{equation}
Pe^{fading}=\frac{1}{\pi} \int_{0}^{\pi / 2} M_{\gamma}\left(-\frac{\mathcal{V}^{2}}{2 \sin ^{2} \omega}\right) d  \omega
\end{equation}
 $\mathcal{V}$ is a commonality constant, depends on the modulation scheme [14, pp 101 (5.1)], In this case, $\mathcal{V}$ depends on the proposed PLNC detection scheme. The approximate AWGN error performance from ({\ref{approxpe}}) is substituted inside the joint MGF at (\ref{jointMGFd}) for the approximate fading Pe and is given by 
\begin{multline}
Pe_{approx}^{fading},R=\\ \frac{1}{\pi}  \int_{0}^{\pi / 2}  \left({1-\frac{N_1(4-\pi) ~min(Ps_1, Ps_2) SNR_1}{8  sin^2 \omega }}\right)^{-0.5} \\ \exp \left(\frac{\frac{ -N_1^{2} \pi ~min(Ps_1, Ps_2) SNR_1}{16 sin^2 \omega}}{1-\frac{ N_1(4-\pi) ~min(Ps_1, Ps_2) SNR_1}{8 sin^2 \omega}}\right) \times \\ \left({1-\frac{ N_2(4-\pi) { ~min(Ps_1, Ps_2)} SNR_2}{8  sin^2 \omega }}\right)^{-0.5} \\ \exp \left(\frac{\frac{ -N_2^{2} \pi { ~min(Ps_1, Ps_2)} SNR_2}{16 sin^2 \omega}}{1-\frac{ N_2(4-\pi) { ~min(Ps_1, Ps_2)} SNR_2}{8  sin^2 \omega }}\right) d \omega
\label{jointMGFdsin}
\end{multline}
Upper-bound of (\ref{jointMGFdsin}) is found by substituting $\omega=\pi/2$, and is given by
\begin{multline}
Pe_{approx}^{fading},R \leq \frac{1}{2} \left({1-\frac{N_1(4-\pi) ~min(Ps_1, Ps_2) SNR_1}{8}}\right)^{-0.5} \\ \exp \left(\frac{\frac{ -N_1^{2} \pi ~min(Ps_1, Ps_2) SNR_1}{16}}{1-\frac{ N_1(4-\pi) ~min(Ps_1, Ps_2) SNR_1}{8}}\right) \times \\ \left({1-\frac{ N_2(4-\pi) { ~min(Ps_1, Ps_2)} SNR_2}{8}}\right)^{-0.5} \\ \exp \left(\frac{\frac{ -N_2^{2} \pi { ~min(Ps_1, Ps_2)} SNR_2}{16}}{1-\frac{ N_2(4-\pi) { ~min(Ps_1, Ps_2)} SNR_2}{8}}\right) d \omega
\label{jointMGFdsin}
\end{multline}

Since RIS elements works in the negative SNR regions, it is vital to find the exact error performance metrics. The exact BER expression is obtained by substituting ({\ref{exactpe}}) in (\ref{jointMGFd}), the resultant expression is given in Appendix.

In time slot 1, the instantaneous SNRs of direct paths are expressed using  (\ref{eqn1}) and (\ref{eqn2}),
\begin{equation}
\gamma_{S_{i}D_{i}}= \frac{{Ps_1} |\mathbf{h}_{S_iD_i}|^2}{N_0}~~i\in(1,2)
\end{equation}
For the cases, $Pe_{S_1D_1}, Pe_{S_2D_2}$, the generalized upper-bound of BER for binary PSK is given as $Pe_{S_iD_i} \leq$ 
\begin{equation}
 \frac{1}{2} \left(\frac{1}{{1+\frac{(N_{AR})(4-\pi) Ps_i}{2 N_0  \sin ^{2} \omega }}}\right)^{\frac{1}{2}} \exp \left(\frac{-\frac{(N_{AR})^{2} \pi Ps_i}{4 N_0 \sin ^{2} \omega }}{1+\frac{(N_{AR})(4-\pi) Ps_i}{2 N_0  \sin ^{2} \omega}}\right) 
\end{equation}
here $N_{AR}$ is the allocated number of RIS reflecting elements for the selected node. Similarly for $Pe_{RD_1}$ and $Pe_{RD_2}$. In time slot2, the instantaneous SNR at destination nodes is given as 
\begin{equation}
\gamma_{RD_i}= \frac{{P_R} |\mathbf{h}_{RD_i}|^2}{N_0} ~~i\in(1,2)
\end{equation}
 BPSK BER is given as $Pe_{RD_i} \leq$ 
\begin{equation}
\frac{1}{2} \left(\frac{1}{1+\frac{(N_{AD})(4-\pi) P_R}{2 N_0  \sin ^{2} \omega }}\right)^{\frac{1}{2}} \exp \left(\frac{-\frac{(N_{AD})^{2} \pi  P_R}{4 N_0 \sin ^{2} \omega }}{1+\frac{(N_{AD})(4-\pi)  P_R}{2 N_0  \sin ^{2} \omega}}\right) 
\end{equation} 
Overall BER at destinations is given by 
\begin{equation}
Pe_{D_i}^{Overall}=Pe_{S_iD_i} + (Pe_R ) (Pe_{RD_i}) ~~i\in(1,2)
\end{equation}

\section{Results and Discussions}
\begin{figure}
\centering
\includegraphics[width=\linewidth]{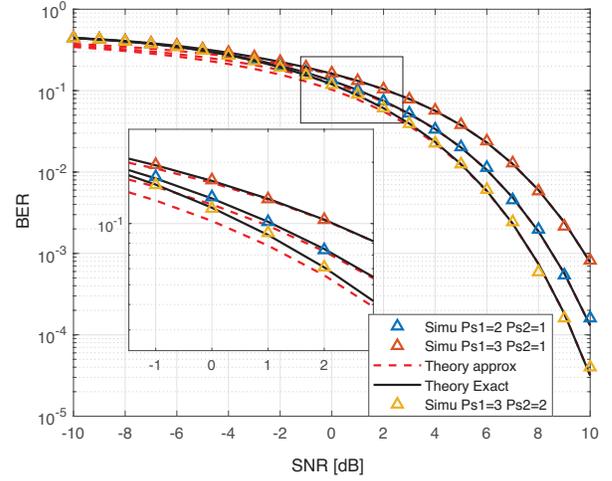}
\caption{BER performance at relay Node for AWGN Environment}
\label{awgnerrper}
\end{figure}
\begin{figure}
\centering
\includegraphics[width=\linewidth]{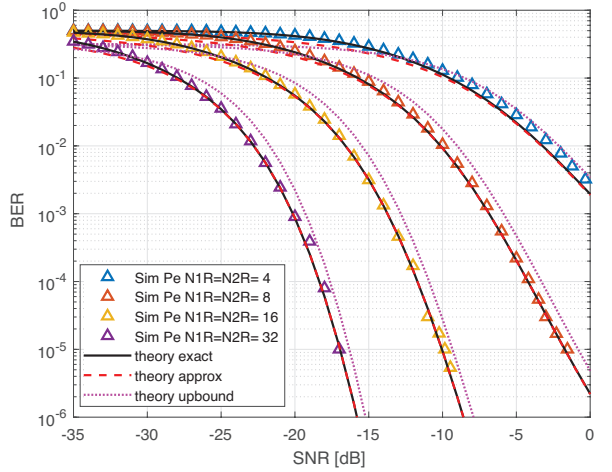}
\caption{BER Performance at Relay Node for Fading Environment $Ps_1$=2 and $Ps_2$=1}
\label{drelayber}
\end{figure}
\begin{figure}
\centering
\includegraphics[width=\linewidth]{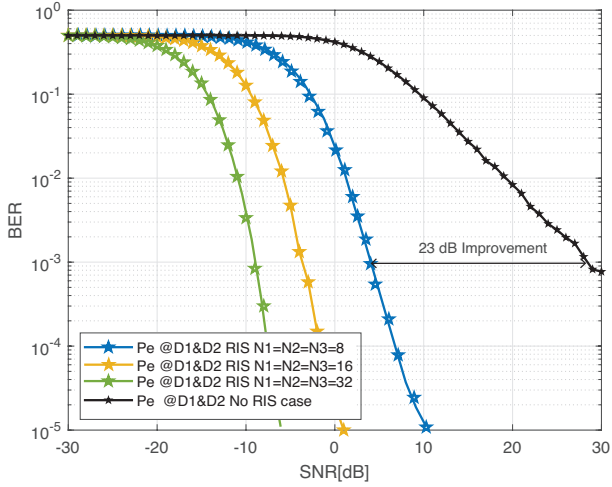}
\caption{End to End BER Performance at Destinations $D_1$ and $D_2$ with $Ps_1$=2 and $Ps_2$=1 }
\label{overall}
\end{figure}
At time slot 1, an optimal Joint Maximum Likelihood detector is employed at relay node $R$. Mathematically, it is expressed as $\underset{x_{1}, x_{2} \in \mathcal{Z}}{\arg \ ~min} \left\|y_{R}-\sqrt{Ps_1}\left[\mathbf{h}_{S_1R}\Phi_2\right]  x_{1}-\sqrt{Ps_2}\left[\mathbf{h}_{S_2R}\Psi_2\right]  x_{2}\right\|^{2}$. where $\mathcal{Z}=\{(-1,-1),(-1,+1),(1,-1),(1,1),\}$.

Exact and approximate BER performances at relay node $R$ in AWGN environment are illustrated in Fig. \ref{awgnerrper} using the expressions (\ref{exactpe}) and (\ref{approxpe}) respectively. It is inferred that the BER performance matches with the proposed duobinary like PLNC detection.

BER performance at relay node $R$ for fading environment is depicted in Fig. \ref{drelayber} using (\ref{jointMGFdsinexact}). For the SNR regions of interest, tight match is observed between simulation and exact theoretical results. Upper-bound of the error probability is also plotted to provide better understanding of reliability of the system.

The end to end BER performance is shown in Fig. \ref{overall}. To achieve error performance metric of $10^{-3}$, conventional PLNC system would require $28 dB$ SNR but the proposed RIS assisted multicast network case with $N_1=N_2=N_3=8$ elements (i.e., 4 elements for direct path, 4 elements for relay path and 4 element each for $D_1$ and $D_2$) requires only $5dB$ ($23 dB$ improvement compared to no RIS case). For $N_1=N_2=N_3=16$ element case and $N_1=N_2=N_3=32$ requires  $-4 dB$ and $-9 dB$ respectively.

\section{Conclusion}
A futuristic RIS assisted multicast communication network is proposed in this paper. Analytical expressions are derived for BER at relay node and End to End BER at destination nodes. The BER performance is compared with conventional multicast relay network. The proposed network outperforms the conventional multicast communication network in terms of reliability and spectrum utilization. This proposed multicast network can be used in applications such as Ultra Reliable Low Latency Communications (URLLC) and Internet of Devices (IoD) swarm networks.


\section{Appendix}
The exact BER expression for fading environment at relay node is given as :
\begin{strip}
\begin{multline}
Pe_{exact}^{fading}=   \frac{1}{\pi}  \int_{0}^{\pi / 2} \left[ \sum_{i=1}^3 (-1)^{(i+1)}  \left({1-\frac{N_1(4-\pi) [Pg_i] SNR_1}{8  sin^2 \omega }}\right)^{-0.5}  \exp \left(\frac{\frac{ -N_1^{2} \pi [Pg_i] SNR_1}{16 sin^2 \omega}}{1-\frac{ N_1(4-\pi) [Pg_i] SNR_1}{8 sin^2 \omega}}\right) \times \right. \\ \left.  \left({1-\frac{ N_2(4-\pi) { [Pg_i]} SNR_2}{8  sin^2 \omega }}\right)^{-0.5}  \exp \left(\frac{\frac{ -N_2^{2} \pi { [Pg_i]} SNR_2}{16 sin^2 \omega}}{1-\frac{ N_2(4-\pi) { [Pg_i]} SNR_2}{8  sin^2 \omega }}\right) \right]
\label{jointMGFdsinexact}
\end{multline} 
\end{strip} 
\begin{strip}
where $Pg_i$ is the generalized power factor and $Pg_1= min(Ps_1, Ps_2)$, $Pg_2=  \left(2\sqrt{Ps_1}+\sqrt{{ ~min(Ps_1, Ps_2)}}\right)^2$, $Pg_3=  \left(2\sqrt{Ps_1}-\sqrt{{ ~min(Ps_1, Ps_2)}}\right)^2$. By substituting $\omega=\pi/2$ in (\ref{jointMGFdsinexact}) upper-bound for the same can be found but is omitted for brevity.
\end{strip}
\end{document}